\documentclass{article}\usepackage{emulateapj}
\usepackage{apjfonts}
\usepackage{psfig}

\usepackage{epsfig,rotating,amsmath}
\slugcomment{Accepted by \it The Astrophysical Journal Letters\rm}

\received{2004 Jan}

\def\gtrapprox{\;\lower 0.5ex\hbox{$\buildrel >\over \sim\ $}}
\def\lessapprox{\;\lower 0.5ex\hbox{$\buildrel < \over \sim\ $}}

\def\Msun  {${\rm M}_\odot$}
\def\deg   {$^\circ$}

\def\HI    {H{$\rm\scriptstyle I$}}

\def\kms   {\ km s$^{-1}$}
\def\Mhi   {M$_{\rm HI}$}

\begin{document}
\title{The Gaseous Trail of the Sagittarius Dwarf Galaxy}

\author{M. E. Putman\altaffilmark{1,2}, C. Thom\altaffilmark{3}, B.K. Gibson\altaffilmark{3}, L. Staveley-Smith\altaffilmark{4}}

\altaffiltext{1}{Center for Astrophysics and Space Astronomy, University of Colorado, Boulder, CO 80309-0389; mputman@casa.colorado.edu} 
\altaffiltext{2}{Hubble Fellow}
\altaffiltext{3}{Centre for Astrophysics \& Supercomputing,
                 Swinburne University, Mail \#31, P.O. Box 218,
                 Hawthorn, VIC, Australia 3122; bgibson@astro.swin.edu.au, cthom@astro.swin.edu.au}
\altaffiltext{4}{Australia Telescope National Facility, CSIRO, P.O. Box 76, Epping, NSW 1710 Australia; Lister.Staveley-Smith@csiro.au}

\begin{abstract} A possible gaseous component to the stream of debris from the Sagittarius 
dwarf galaxy is presented.  We identify $4 - 10 \times 10^6$ \Msun\ of neutral
hydrogen along
the orbit of the Sgr dwarf in the direction of the Galactic anticenter
(at 36 kpc, the distance to the stellar debris in this region).
This is 1-2\% of the estimated total mass of the Sgr dwarf.
Both the stellar and gaseous components have negative velocities, 
but the gaseous component extends to higher
negative velocities.
If associated, this gaseous stream was most likely stripped from the main body of the dwarf 
0.2 - 0.3 Gyr ago during its
current orbit after a passage through a diffuse edge of the Galactic
disk with a density $> 10^{-4}$ cm$^{-3}$.  This gas represents the dwarf's 
last source of star formation fuel and explains how the galaxy was forming
stars 0.5-2 Gyr ago.

\end{abstract}


\section{Introduction}

Evidence for the process which formed our Galaxy is found throughout
our Galaxy's halo as trails of stars and gas (e.g. Yanny et al. 2003;
Putman et al. 2003; hereafter P03).  These Galactic building blocks are currently
accreting satellites, and the Sagittarius dwarf galaxy
(hereafter Sgr dwarf) is one of the closest examples of this process (Ibata et
al. 1994).  The evidence continues to accumulate that contiguous 
streams of leading and trailing stellar debris are being pulled from
the Sgr dwarf as it spirals into the Milky Way
(e.g. Newberg et al. 2003; Majewski et al. 2003; hereafter M03).  The
stars associated with the Sgr dwarf span a wide range of ages, with the youngest
population between 0.5 - 2 Gyr old (Layden \& Sarajedini 2000; Dolphin
2002; M03).  This indicates that within the past Gyr, the Sgr dwarf
was forming stars and had a source of star formation fuel.

Neutral hydrogen is a principal source of star formation
fuel for a galaxy.  Galaxies which contain HI are commonly currently 
forming stars (e.g., Lee et al. 2002; Meurer et
al. 2003) and those without detectable HI tend to have primarily an older stellar
population and thus appear to have exhausted their star formation
fuel (e.g., Gavazzi et al. 2002).  This is evident in the dwarf
galaxies of the Local Group.  The stars in the
Local Group dwarfs vary from being almost entirely ancient ($>$ 10 Gyr;
e.g., Ursa Minor) to a number of systems which are actively forming stars
(e.g. WLM, Phoenix, LMC).  The HI content of the dwarfs
is summarized by Mateo (1998), Grebel, Gallagher, \& Harbeck (2003), 
and Bouchard et al. (2003).
The majority of the Local Group galaxies
with HI have formed stars within the past 2 Gyr and those
with no evidence for recent star formation do not contain detectable HI (e.g., Mateo
1998; Dolphin 2002).

Pointed HI observations of the central region of the Sgr dwarf
($\alpha$, $\delta$ = 19$^{h}$ 00$^{m}$, -30\deg\ 25\arcmin\ (J2000); $l,b = $6\deg,
-15\deg) indicate that our closest satellite galaxy does not currently
contain a significant amount of star formation fuel (\Mhi\ $< 1.5
\times 10^{4}$ \Msun\ (3$\sigma$); Koribalski, Johnston \& Otrupcek
1994).  The search for HI associated with the Sgr dwarf was continued 
by Burton \& Lockman (1999); but they also found no associated gas over 18
deg$^2$ between $b = -13$\deg\ to -18.5\deg\ with limits of \Mhi\ $< 7 \times
10^{3}$ \Msun\ (3$\sigma$).  These results are surprising considering
the Sgr dwarf was forming stars within the last Gyr.  Since the
orbit of the Sgr dwarf is approximately 0.7 Gyr (Ibata \& Lewis 1998), one might expect to find this fuel stripped
along the dwarfs orbit, possibly at a similar location to the stellar
trail.  The trailing stellar tidal tail of the Sgr dwarf has
recently been found to extend for over 150\deg\ across the South
Galactic Hemisphere with a mean distance between 20 to 40 kpc from the
Sun (M03).  Here we present HI data from HIPASS
(HI Parkes All-Sky Survey\footnote{The Parkes Telescope is part of the Australia Telescope 
which is funded by the Commonwealth of Australia for operation as a National 
Facility managed by CSIRO.}) along the entire Sgr dwarf galaxy orbit to
investigate the possibility that a gaseous Sgr trail is also present.
The gas detected represents a potential method of tracing the
history, make-up, and classification of the Sgr dwarf galaxy, as well
as the construction of our Galaxy.

\section{Observations}

The neutral hydrogen data are from the \HI\ Parkes All--Sky Survey
(HIPASS) reduced with the {\sc minmed5} method (P03).
HIPASS is a survey for \HI\ in the Southern sky, extending from the
South celestial pole to Decl. $= +25$\deg, over velocities from $-1280$
to $+12700$ km s$^{-1}$ (Barnes  et al. 2001).  
The survey utilized the 64--m Parkes radio telescope,
with a focal--plane array of 13 beams arranged in a hexagonal grid, to
scan the sky in $8^\circ$ zones of Decl. with Nyquist sampling.  The
%spectrometer has 1024 channels for each polarisation and beam, with a
%velocity spacing of 13.2 km s$^{-1}$ between channels 
{\sc minmed5} reduced HIPASS data has a spatial resolution of
15.5\arcmin\ and a spectral resolution, after Hanning smoothing, 
of 26.4 km s$^{-1}$.   The
survey was completed with a repetitive scanning procedure which
provides source confirmation, mitigates diurnal
influences, and aids interference excision.
%The data was reduced with the {\sc minmed5} method which was designed to recover
%extended emission in the HIPASS data for the purpose of studying
%HVCs (see P03 for a full description of the reduction and gridding).  
%This procedure greatly increases the sensitivity of the data to
%large--scale structure without substantial loss of flux density,
%except when the emission fills the entire 8\deg~ scan (i.e. the Galactic
%Plane).  \HI\ emission in  the {\sc lsr} velocity range
%$-700$ to $+1000$ \kms~was reduced in this manner and the calibrated
%scans were gridded with the median method described by Barnes et
%al.~(2001), without the weighting which overcorrects the fluxes for
%extended sources. 
For extended sources, the RMS noise is 10 mJy
beam$^{-1}$ (beam area 243 arcmin$^2$), corresponding to a brightness
temperature sensitivity of 8 mK. 
The northern extension of the survey, from $+2$\deg\
to $+25$\deg, was only recently completed and is presented for
the first time here.  The noise in these cubes is slightly elevated
compared to the southern data (11 mK vs. 8 mK).  This may be
due to a combination of low zenith angles during these
observations and an inability to avoid solar interference as
effectively.   
%The spatial size of the gridded
%cubes is 24\deg~ $\times$ 24\deg~ with a few degrees of overlap
%between each cube.  
Integrated intensity maps of the high positive and
negative velocity gas (generally $|$v$_{\rm lsr}| > 80$ \kms, as long as the gas was clearly separate from
Galactic emission) were made for the 24 deg$^2$ cubes which lie along the Sgr
orbit.  The positive velocity cubes had very little emission in
them, so we concentrated on the negative velocity cubes.
The noise at the edges of the negative velocity maps were blanked within
{\sc aips} and the images were then read into {\sc idl} to create the map of 
the entire orbit shown in Figure 1. 
At 20-40 kpc the {\sc minmed5} reduced HIPASS data has a sensitivity to 
clouds of gas with \Mhi\ $> 80-320$ \Msun\ ($\Delta$v = 25 \kms; 3$\sigma$).

\section{Results}

The large scale HI map which includes all of
the high negative velocity gas along the orbit of the Sgr dwarf
is shown in Figure 1.  
This plot is in Celestial coordinates as it depicts the main features
found along the orbit better than Galactic coordinates.  The Galactic Plane, 
%is shown by the solid line with the dotted lines on each side representing
%$+/-$ 5\deg\ from $b=0$.  
the orbit of the Sgr dwarf (Ibata \& Lewis 1998), 
%is shown by the solid line extending across the plot horizontally, 
and the current position of the Sgr dwarf are shown.
The stream of M giants presented by M03 has quite a broad width (commonly
several degrees) along this orbit.
 The orbit of the Sgr dwarf crosses the Galactic Plane at $\ell \sim 185$\deg\ and $\ell \sim
0$\deg\ which corresponds to approximately ($\alpha, \delta$)$ = 6^{h}$, 25\deg\ 
and 18$^{h}$, -30\deg, respectively.  At the Galactic Center high negative and positive velocity
gas is present, thus we have labeled the negative velocity HI emission within $+/-$ 5\deg\ of the Galactic
plane at that location.  The Sgr orbit crosses the
Magellanic Stream (labeled) at the South Galactic Pole or ($\alpha, \delta$)$ = 0^{h}$, -15\deg.  
The majority of the Sgr dwarf orbit does not have both stars and HI
gas at high negative velocities,  with the
exception of the HI gas between $\alpha = 3 - 4.5$ h and $\delta = 0 -
30$\deg, or $\ell \approx 155 - 195$\deg\ and $b \approx -5 - -50$\deg.  
 A close up of the velocity distribution of these HI clouds in Galactic coordinates is shown in the channel maps of Figure 2. 
%This plot shows the velocity distribution of the high-velocity gas with colors, as well
%as column density contours.  
The M giants of the Sgr stellar stream fill a large percentage
of the region shown Fig. 2.  This high velocity HI gas was previously
identified as part of the Anticenter complexes (ACHV and ACVHV; Wakker \& van Woerden
1991) and was discovered almost 40 years ago (e.g., Mathewson et al. 1966), but its
relationship to the Sgr dwarf was not previously noted.
%distance constraint probably isn't worth mentioning >350pc for lower velocity gas
There are some small negative velocity clouds along the Sgr dwarf orbit, but besides
the labeled gas in Fig. 1,
the only other substantial complex of negative velocity gas along the orbit of the Sgr dwarf 
is in a region which has predominantly positive velocity stars (Ibata et al. 2001).  This
is the group of high-velocity clouds known as Complex L at 
$\alpha \approx $ 15.25 h, $\delta \approx $ -19.5\deg. 
The amount of high positive velocity gas along the orbit of the Sgr
dwarf is very limited, and not correlated with the position of the
positive velocity stars along the Sgr stellar stream. 

Figure 2 shows the gas along the orbit of the Sgr dwarf 
between $\alpha = 2 - 5.5$ h and $\delta = 0 - 30$\deg\ 
in Galactic coordinates over the velocity range 
of -380 to -85 \kms\ (LSR).  The emission extending from -380 
to -180 \kms\ at $b < -20$\deg\
is orientated along the orbit of the Sgr dwarf and is completely isolated
from gas which merges with lower velocity emission.  
This complex of gas would amount to M$_{HI} = 4.3 \times
10^6$ \Msun\ at 36 kpc, the approximate distance to
the stars in the Sgr stream.
Other emission which
appears in the channel maps include a distinct  
filament beginning at $b = -10$\deg\ in the velocity range of
-245 to -75 \kms\ that would
have a mass of $5.4 \times 10^6$ \Msun\ at 36 kpc. 
The gas in both complexes has peak column densities on the order of $10^{20}$ cm$^{-2}$ at the
15.5\arcmin\ resolution of HIPASS and extends to the column density
limits of the data ($5\sigma \sim 3 \times 10^{18}$ cm$^{-2}$; $\Delta$v $= 25$ \kms).
 There is also
a population of clouds appearing at -125 \kms\ that merge
into intermediate velocity emission and are not included in Figure 1
for this reason.  The distance of 36 kpc was adopted based on the continuous
stellar trail presented in M03 at those distances and the matching
distance of a population of carbon stars (Ibata et al. 2001).  The carbon stars
associated with the Sgr dwarf at this position have velocities between 
-140 to -160 \kms\ (LSR; Dinescu et al. 2002; Totten \&
Irwin 1998; Green et al. 1994).  

%Ibata et al. (2001) also associate a stellar component at $\sim$15 kpc in this
%direction with the Sgr dwarf, but these stars do not have matching velocities to
%the HI gas presented here.

\section{Discussion}

The present orbit of the Sgr
dwarf is estimated to be 0.7 Gyr (Ibata \& Lewis 1998).   Using this
orbit, the core of the Sgr dwarf was at the position of the HI complex
presented here approximately 0.2 - 0.3 Gyr ago.  It would make sense if 
the gas was part of the Sgr dwarf 0.3 - 0.7 Gyr ago considering
the age of part of the Sgr dwarf stellar population (0.5 - 2 Gyr; Layden \& 
Sarajedini 2000; Dolphin 2002; M03).  Not surprisingly star formation
surveys find that those galaxies which are actively forming stars
also contain substantial amounts of star formation fuel in the
form of neutral hydrogen (Lee et al. 2002 (KISS); Meurer et al. 2003 (SINGG)).
This paper addresses the issue of when and how the star formation fuel
of the Sgr dwarf was stripped from the galaxy.
  
The gas between -380 and -180 \kms\ at $b<-20$\deg\ in Figure 2 is
the gas we propose was most likely once part of the Sgr dwarf, as it follows
the orbit of the Sgr dwarf and is completely isolated from lower velocity emission
that may have a relation to disk gas (e.g., Tamanaha 1995).
This gas lies at a position where the Sgr dwarf would have passed
through the extreme edge of the Galactic disk, and at 
this position the Sgr dwarf stellar
debris is approximately 40 $-$ 50 kpc from the Galactic Center (M03).  
We have adopted 45 kpc as the typical distance from the Galactic Center at 
the position of the bulk of the gas.  This distance is significantly beyond the
typical radius quoted for our Galaxy ($\sim$26 kpc), however it
is possible that an extended ionized disk exists for our Galaxy (e.g., Savage
et al. 2003), 
as found in other systems (Maloney 1993; Bland-Hawthorn, Freeman \& Quinn 1997; 
Steidel et al. 2002).
The passage through an extended disk of our Galaxy, in addition to
the tidal forces already obviously at work as evident from the stellar
tidal stream, might have been enough to disrupt the HI in the core of
the galaxy and cause the dwarf to lose all remaining star formation
fuel.  The gas will be stripped from a galaxy if $\rho_{IGM} v^2 > \sigma^2 \rho_{gas} / 3$ (Mori \& Burkert 2000; Gunn \& Gott 1972).
We can use this equation to 
estimate the density needed at the edge of the disk to strip the gas
from the core of the Sgr dwarf.  We use a
tangential velocity of 280 \kms\ for the Sgr dwarf and a velocity
dispersion of 11.4 \kms\ (Ibata et al. 2003; Ibata \& Lewis 1998).  If the column
densities and size of the HI distribution in the core of the Sgr dwarf
were on the order of $5 \times 10^{20}$ cm$^{-2}$ (averaged over the core)
and 1 kpc, the typical $\rho_{gas}$ is 0.16 cm$^{-3}$.
An extended disk density greater than $3 \times 10^{-4}$ cm$^{-3}$ is 
then needed to strip the gas via ram pressure stripping.  Based on 
previous estimates of the density of the Galactic
halo, this density would be
easily achieved in the plane of Galaxy, 20 kpc from the currently
observed edge of the disk.  Examples of estimates for
the density of the Galactic halo which are consistent with $\sim 10^{-4}$ cm$^{-3}$ 
at distances of 50 kpc and more include:  explaining the existence of head-tail clouds which can be associated
with the Magellanic Stream (Quilis \& Moore 2001); the
confinement of the tip of the Magellanic Stream (Stanimirovic et al. 2002);
the O VI high-velocity absorption line results (Sembach et al. 2003); and the diffuse x-ray 
emission (Wang \& McCray 1993) and dispersion measures for LMC pulsars (Taylor, Manchester \& Lyne 1993)
when the ionized medium is extended over $\sim$100 kpc.

Stripping gas from a Galactic satellite by passing it through
the edge of the Galaxy's disk was proposed as a possible
solution for the origin of the Magellanic Stream by Moore \& 
Davis (1994; hereafter MD94).
They invoked an extended ionized disk at 65 kpc from
the Galactic Center with column densities $< 10^{19}$ cm$^{-2}$
to strip gas from the satellite galaxy.
%with column densities between $10^{19}$ to $10^{20}$ cm$^{-2}$ 
The stripped gas initially has its speed reduced, but the decelerated
gas falls to a lower orbit and subsequently attains a higher
angular velocity. 
% by $\sim$25-130 \kms\ over 0.5 Gyr after 
In order to reproduce the properties and distribution of the Magellanic Stream, MD94 
induce a braking effect on the gas with a diffuse
medium in the halo, preventing it from rapidly falling into the Galactic potential.
%reaching very high negative velocities, and appearing to lead or overlap with
%the Magellanic Clouds. 
Though the HI trail of the Sgr dwarf has a very different
structure from the Magellanic Stream, parallels can be made between 
the MD94 model and the stripping of the Sgr HI clouds.
Both of the systems have similar impact velocities and have gas at similar
distances from their cores (roughly 50-60 kpc) with similar column densities
and velocities.  The gas along the Sgr dwarf orbit has slightly higher negative 
velocities than the carbon stars with velocity determinations in this region
and this may be due to the gas falling into a slightly lower
orbit than the stars.  The exact predicted position and velocity of the gas relative to
the stars will be addressed in a future simulation paper.
%Gas in the Magellanic Stream with column densities similar
%to the gas here has velocities of -100 to -250 \kms.

At 36 kpc from the sun the HI gas between -380 and -180 \kms\ at $b<-20$\deg\
has a mass of $4.3 \times 10^6$ \Msun.
%This gas is at more negative velocities than the carbon stars with velocity determinations
%in this region ($\sim$ -140 to -160 \kms), but this is not unexpected with the
%addition of ram pressure forces to the tidal.
If the filament extending from $b= -10$ to $-25$\deg\ and v$_{\rm LSR} = -245$ to $-85$ \kms\
is included as once being part of the Sgr dwarf, the total HI mass
goes up to $9.7 \times 10^6$ \Msun.
A typical total HI mass 
for a dwarf galaxy is on the order of $10^{7-8}$ \Msun\ (Grebel et al. 2003),
so $10^{6-7}$ \Msun\ of gas originally being associated with the dwarf is
certainly plausible and may represent anywhere from the majority to 10\%
of the Sgr dwarf's original HI mass.  Using
a total Sgr dwarf mass of $5 \times 10^8$ \Msun\ (M03), this
gas represents 1-2\% of its total mass.
The Sgr dwarf most likely originally had more than $10^{6-7}$ \Msun\ of
neutral hydrogen associated with it, as the outer gaseous component would have 
been the first thing stripped (e.g., Mihos 2001; Yoshizawa \&
 Noguchi 2003).
% and evident in the interaction of the Magellanic Clouds with the Milky
%Way which exhibits a gaseous stream with no stellar counterpart (P03).  
This outer gas would have had column densities between
$10^{18} - 10^{19}$ cm$^{-2}$ and has most likely either already
dispersed or been ionized, although remnants of this
gas may be present as small HVCs along the Sgr dwarf orbit.   Since there is
currently no HI associated with the core of the Sgr dwarf (Koribalski
et al. 1994), and the dwarf has stars which are 0.5 - 2 Gyr
old, the HI gas 
presented here most likely represents the 
high column density gas from the core of the Sgr dwarf which was finally stripped
when the dwarf passed through the higher density medium in the
extended Galactic disk.  This is 
supported by the relatively high peak column densities
of the HI gas currently ($\sim 10^{20}$ cm$^{-2}$). 
For the Sgr HI gas to survive for 0.2-0.3 Gyr (the time since
the last passage of the core Sgr dwarf), not to mention the
Magellanic Stream which is thought to be $>$ 0.5 Gyr old, the gas should either be
confined by an existing halo medium or associated with significant
amounts of dark matter.

\section{Overview}

Our Galaxy's halo is made up of numerous streams of satellite debris which trace
its formation.
The association of HI gas with the Sagittarius dwarf galaxy emphasizes
the number of dynamical processes occuring within the Galactic halo.  
In this paper we find $4 - 10 \times 10^6$ \Msun\ of HI gas along
the orbit of the Sgr dwarf.
% at similar velocities to the Sgr stellar stream in this direction.  
This is approximately 1-2\% of the total mass of the dwarf, 10-20 
times lower than the HI mass of the Magellanic Stream, and
approximately the same HI mass as the Leading Arm of the Magellanic System (P03).  
We argue this HI was the last gas stripped from the core of the Sgr
dwarf $\sim 0.2$ - 0.3 Gyr ago due to its passage through the extended Galactic 
disk with densities $> 10^{-4}$ cm$^{-3}$.  This is supported by 
the agreement in the spatial distribution of
the gas and stars along the Sgr orbit, both components having high
negative velocities, the location of the gas relative to 
the plane of our Galaxy, the relatively 
high column densities of the HI gas, and the star formation history of
the Sgr dwarf.   The association of HI gas with the Sgr
stellar stream suggests that the dwarf spheroidal classification for
the Sagittarius galaxy may need to be reconsidered.
It also suggests that slightly offset HI and stellar streams may be a
common feature of disrupted satellites in the Galactic halo.  With
the accretion of this Sgr HI stream, the Magellanic Stream (P03),
Complex C (Wakker et al. 1999), and possibly other HVCs, there is ample
fuel for our Galaxy's continuing star formation and understanding the
distribution of stellar metallicities (e.g., the G-dwarf problem is not a problem). 
Determining the metallicity, distance, and ionization properties of the HI
gas presented here will aid in confirming if this HI gas was indeed stripped from
the Sgr dwarf during its current orbit.  Since we propose this is the last gas
from the core of the Sgr dwarf, we will also look for molecular gas and
dust associated with these clouds.   A future simulation paper will address
the exact spatial and velocity distribution expected for gas stripped 
from the Sgr dwarf.

\acknowledgements{We would like to thank Phil Maloney, Steve
Majewski and Geraint Lewis for useful discussions.  We also thank
the referee for many useful comments.  M.E.P. acknowledges support by
NASA through Hubble Fellowship grant HST-HF-01132.01 awarded by the Space Telescope Science
Institute, which is operated by AURA Inc. under NASA
contract NAS 5-26555.  B.K.G. acknowledges the support of the Australian Research
Council, through its Large Research Grant and Discovery Project schemes.}

\clearpage

%----------------------------------------------------------------------------------------
\centerline{\psfig{figure=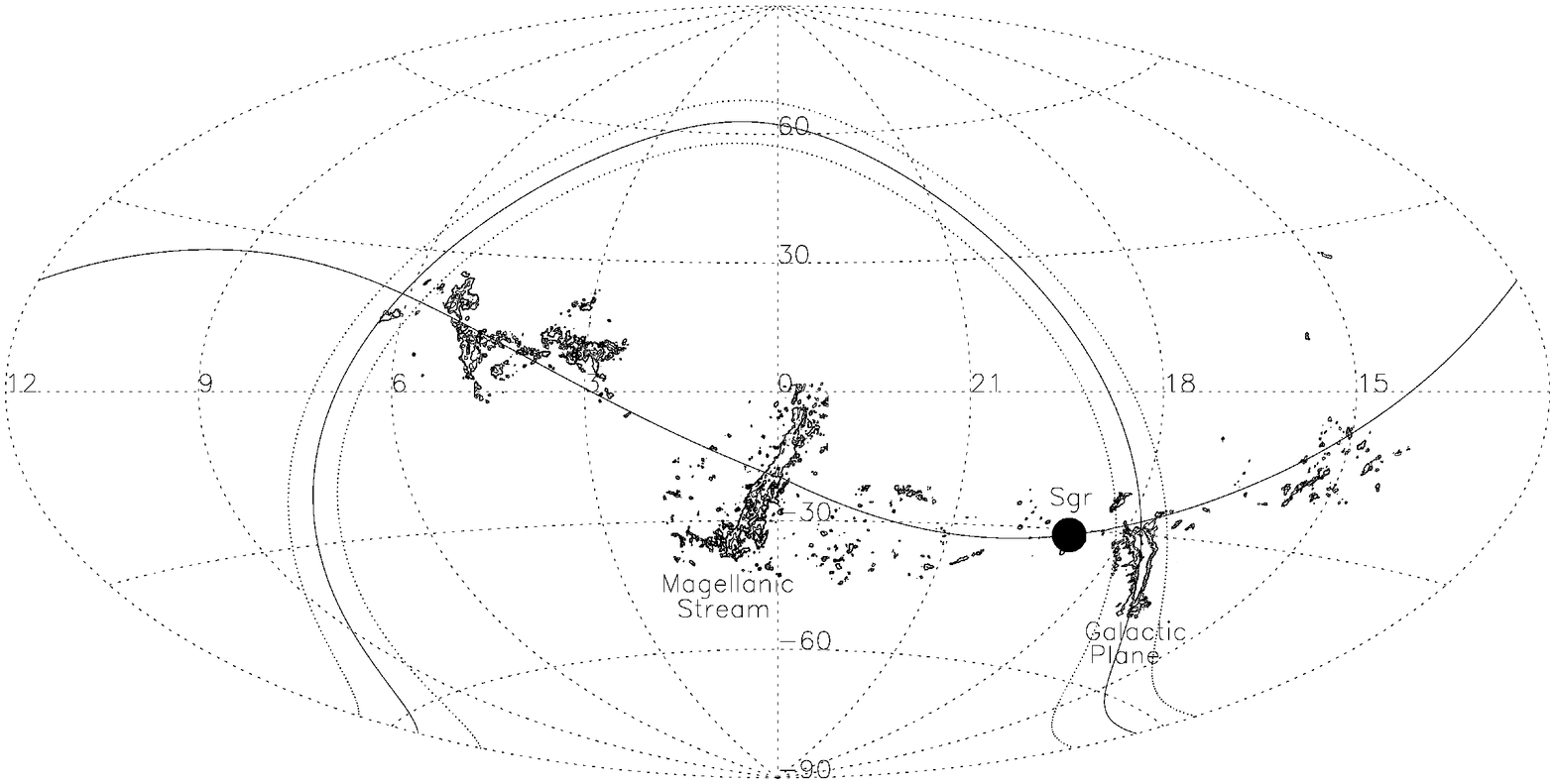,width=10cm,angle=-90}}

\noindent {\sl Fig.~1.} The negative high-velocity HI ($\sim -85$ to $-400 $\kms; LSR) 
along the orbit of the Sgr dwarf galaxy in Celestial coordinates.  
 The current position of the Sgr dwarf is shown by the solid point, and the orbit 
of the Sgr dwarf (Ibata \& Lewis 1998) is plotted as the solid line through
this point.
The negative velocity gas attributed to the Magellanic Stream and Galactic Plane
gas at the Galactic Center is 
labeled.  $b = 0$\deg\ is indicated by the solid line through the labeled Galactic
Plane gas, with the two dotted lines
on each side representing $b = +5$\deg\ and $-5$\deg. The negative velocity carbon
stars extend from approximately $\alpha, \delta = 0^{h}, -20$\deg\ to $12^{h}, 20$\deg\
 along the Sgr orbit (Ibata et al. 2001).
Contours represent column density levels of 0.5, 1.0, 5.0, and 10.0 $\times 10^{19}$ cm$^{-2}$. 
%----------------------------------------------------------------------------------------

\medskip

\centerline{\psfig{figure=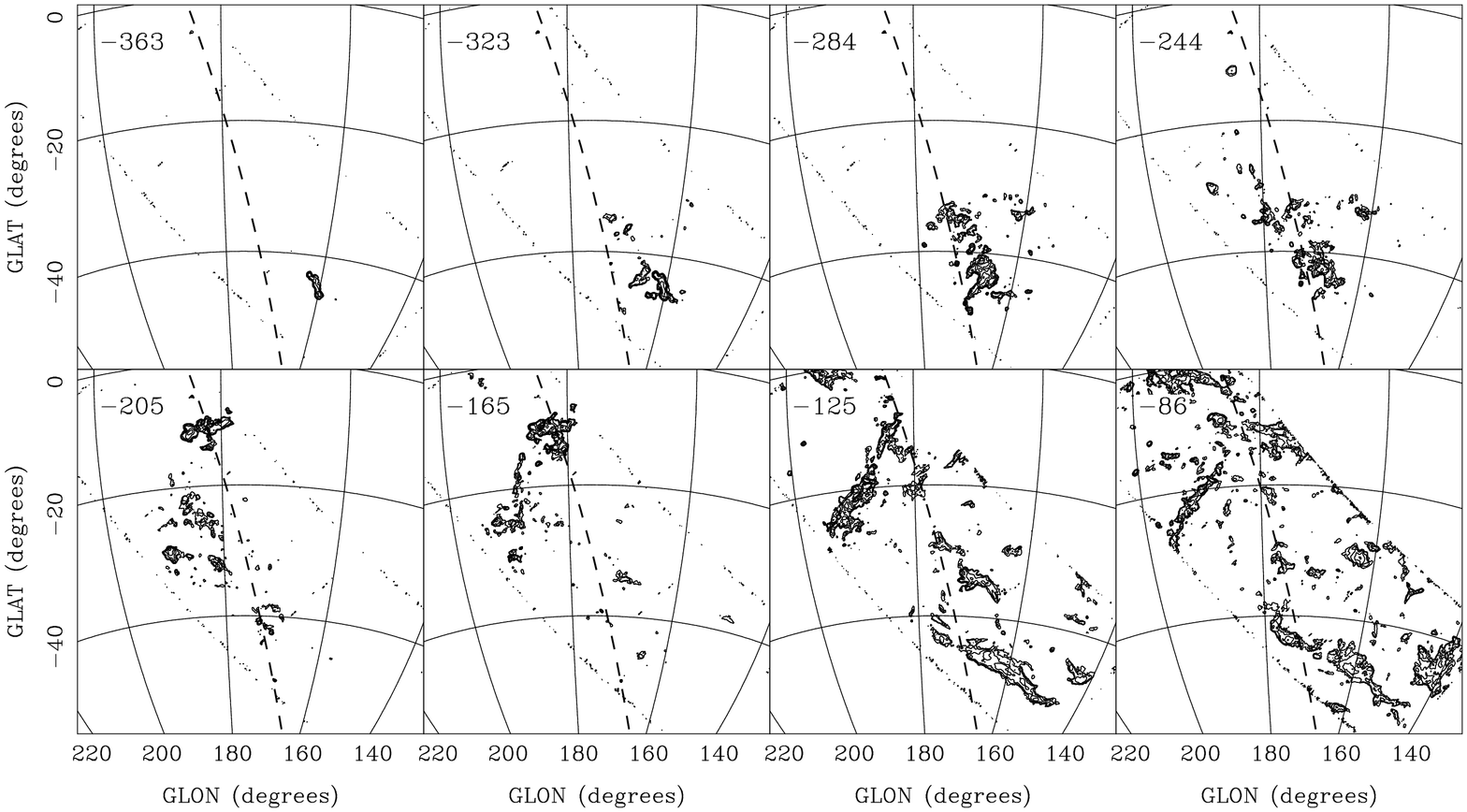,width=15cm,angle=-90}}
\noindent {\sl Fig.~2.}  Channel maps of the HI clouds found along the trailing
stellar stream of the Sgr dwarf in Galactic coordinates with v$_{\rm LSR}$ 
labeled in the upper left corner.  The M giants (M03) extend
across this region in a $\sim$15\deg\ band along the orbit of the Sgr
dwarf (dashed line;
%from $\ell, b = 185$\deg, 0\deg\ to the bottom of the plot at $\ell = 170$\deg;
Ibata \& Lewis 1998).   Contours are 0.055, 0.11, 0.22, 0.44, 0.88, and 1.8 K.

%-------------------------------------------------------------------------------
%\centerline{\psfig{figure=f2.ps,width=14cm,angle=-90}}
%
%\noindent {\sl Fig.~2.} The negative velocity HI clouds found along the trailing stellar stream of the Sgr dwarf at similar velocities to the stars.  The M giants (M03) extend
%across this plot in a $\sim$15\deg\ band along the orbit of the Sgr
%dwarf (dashed line; Ibata et al. 2001).
%%from $\ell, b = 185$\deg, 0\deg\ to the bottom of the plot at $\ell = 170$\deg. 
%Dark blue represents -125 \kms\ (LSR) and
%bright pink represents -370 \kms. Carbon stars in the Sgr stellar stream in this
%region have velocities which range from $-140$ to $-160$ \kms. Contours are
%0.7, 1.4, 2.8, 5.6, and 11.2 $\times 10^{19}$ cm$^{-2}$.

%-------------------------------------------------------------------------------

\end{document}